\documentclass[12pt]{article}
\usepackage{graphicx}

\newcommand\pubnumber{DPF2015-284}
\newcommand\pubdate{\today}

\def\napoli{Department of Physics\\
University of Minnesota, Minneapolis, MN, USA}

\def\Title#1{\begin{center} {\Large #1 } \end{center}}
\def\Author#1{\begin{center}{ \sc #1} \end{center}}
\def\Address#1{\begin{center}{ \it #1} \end{center}}

\newcommand\pubblock{\rightline{\begin{tabular}{l} \pubnumber\\
         \pubdate  \end{tabular}}}
\newenvironment{Abstract}{\begin{quotation}  }{\end{quotation}}
\newenvironment{Presented}{\begin{quotation} \begin{center} 
             PRESENTED AT\end{center}\bigskip 
      \begin{center}\begin{large}}{\end{large}\end{center} \end{quotation}}
\def\Acknowledgments{\bigskip  \bigskip \begin{center} \begin{large}
             \bf ACKNOWLEDGMENTS \end{large}\end{center}}

\def\beq{\begin{equation}}
\def\eeq#1{\label{#1}\end{equation}}
\def\eeqn{\end{equation}}

\def\beqa{\begin{eqnarray}}
\def\eeqa#1{\label{#1}\end{eqnarray}}
\def\eeqan{\end{eqnarray}}

\def\overbar#1{\overline{#1}}

\let\bar=\overbar

\def\Dslash{\not{\hbox{\kern-4pt $D$}}}
\def\dslash{\not{\hbox{\kern-2pt $\del$}}}

\def\msb{{\bar{\ssstyle M \kern -1pt S}}}

\begin{document}
\begin{titlepage}
\pubblock

\vfill
\Title{Search for exotic transitions of muon neutrinos to electron neutrinos with MINOS}
\vfill
\Author{ Marianna Gabrielyan }
 \centerline {(for the MINOS and MINOS+ Collaborations)}
\Address{\napoli}
\vfill
\begin{Abstract}
The observed neutrino flavor transitions are currently explained by the three flavor neutrino oscillation phenomenon, considered to be the leading order mechanism behind the flavor transitions.  Currently existing data from LSND, MiniBooNE  and reactor experiments demonstrate anomalies that could potentially be indications of non-standard neutrino phenomena.  MINOS can probe transitions from muon neutrinos to electron neutrinos and search for anomalous behavior that cannot be explained by standard model neutrino oscillations.  Here we present the search for second order effects in the flavor transitions by analyzing the MINOS $\nu_\mu \to \nu_e$ channel.

\end{Abstract}
\vfill
\begin{Presented}
DPF 2015\\
The Meeting of the American Physical Society\\
Division of Particles and Fields\\
Ann Arbor, Michigan, August 4--8, 2015\\
\end{Presented}
\vfill
\end{titlepage}
\def\thefootnote{\fnsymbol{footnote}}
\setcounter{footnote}{0}

\section{Introduction}

 One of the motivations for exploring the sterile neutrino oscillations in the  $\nu_\mu \to \nu_e$ appearance channel is the existing tension between the short (SBL) and long (LBL) baseline neutrino experiments.  For example, the Liquid Scintillator Neutrino Detector (LSND) and MiniBooNE experiments report anomalous excess of  $\overbar{\nu_e}$ appearance in a $\overbar{\nu_\mu}$ beam over a short baseline, which cannot be explained by standard three flavor oscillations ~\cite{lsnd, miniboone}. 
 %
\begin{figure}[htb]
\centering
\includegraphics[height=3.0in]{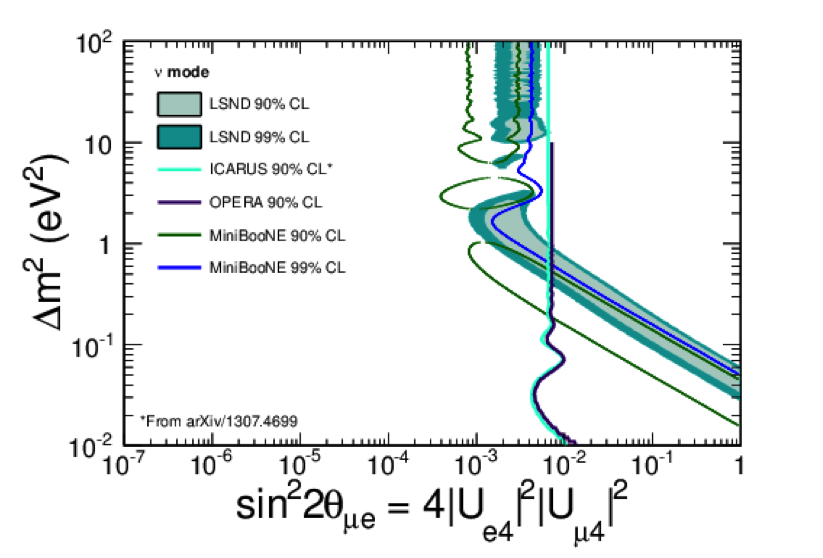}
\caption{LSND and MiniBooNe allowed regions for the sterile mixing angle and mass splitting. Shows tension with the OPERA and ICARUS long baseline experiments. }
\label{fig:LSND_sens}
\end{figure}

 One of the explanations is an existence of one or more sterile neutrino states that are not interacting via the weak interaction. 
Adding a new neutrino type to 3$\nu$ model introduces another mass splitting and three mixing angles to parameter space.  
Figure~\ref{fig:LSND_sens} shows the allowed regions for the additional mass splitting and $\theta_{\mu e}$ mixing angle. 
MINOS+ $\nu_e$ appearance analysis uses (3+1)$\nu$ model to investigate the sterile oscillation hypothesis in the parameter space allowed by LSND and MiniBooNE.

\section{MINOS Experiment}

The Main Injector Neutrino Oscillation Search (MINOS) experiment~\cite{minos} is a long-baseline neutrino oscillation experiment utilizing the Neutrinos at the Main Injector (NuMI) neutrino beam. The experiment consists of two functionally identical tracking calorimeters, primarily optimized to detect $\mu^{+}/\mu^{-}$,  separated by 735 km distance. The main goals of the experiment are to make precise measurements of the neutrino oscillation parameters to other active neutrino flavors, explore the possibility of sterile oscillations as well as to study exotic scenarios such as non-standard neutrino interactions. Near Detector (ND) is located at Fermilab  $\sim1$ km away from NuMI target, is measuring the neutrino flux, cross section, energy spectrum and the flavor composition of the beam.  Far Detector (FD) is located at Soudan Underground Laboratory at Northern Minnesota,  $\sim735$ km away from NuMI target, measures the same beam parameters. Functionally identical design was chosen to effectively cancel out systematic uncertainties related to the neutrino beam flux and neutrino interaction cross sections. 
%
\begin{figure}[htb]
\centering
\includegraphics[height=2.3in]{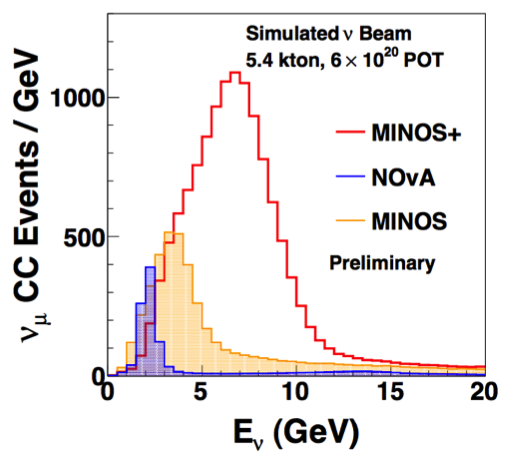}
\caption{Simulated NuMI neutrino energy spectrum for MINOS, MINOS+ and NO$\nu$A experiments. }
\label{fig:E_nu}
\end{figure}
%
 Since the MINOS detectors were not optimized for $\nu_e$ CC events, a new statistical selection 
algorithm was implemented to distinguish $\nu_\mu \to \nu_e$ candidates from background interactions - particularly NC events. $\nu_e$ appearance provides sensitivity to $\theta_{13}$, $\theta_{23}$ octant, mass hierarchy and $\delta_{CP}$.   

MINOS+ is a continuation of the MINOS experiment, operating the same detectors. It will run concurrently with the NO$\nu$A experiment with an upgraded NuMI beam. 
MINOS+ allows access to higher energy region to study sterile neutrino oscillations as well as non-standard neutrino interactions, with beam covering wide range of energies and peaking at 7 GeV as can be seen from Figure~\ref{fig:E_nu}.

\section{NuMI Beamline}

\begin{figure}[]
\centering
\includegraphics[height=1.6in]{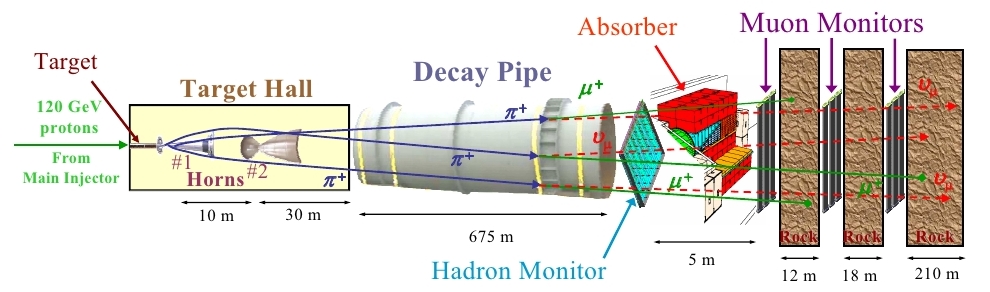}
\caption{Schematic view of NuMI production beamline}
\label{fig:beamline}
\end{figure}

The $\nu_\mu$ neutrino beam is produced by colliding 120 GeV proton beam from the Main Injector with graphite target. Mesons ($\pi^{\pm}/K^{\pm}$), produced as a result of these collisions, are then focused into the decay pipe by focusing magnetic horns, where they are allowed to decay (Figure~\ref{fig:beamline}). 
%
This admixture then passes through an absorber, which stops all charged particles from moving forward. As a result  wide-band on-axis $\nu_\mu$ beam is produced and directed towards the MINOS detectors. By switching the magnetic horn polarity, one can switch from neutrino to antineutrino beam production mode.

\section{Sterile Neutrino Sensitivity in $\nu_e$ Appearance}

\begin{figure}[]
\centering
\includegraphics[height=3.1in]{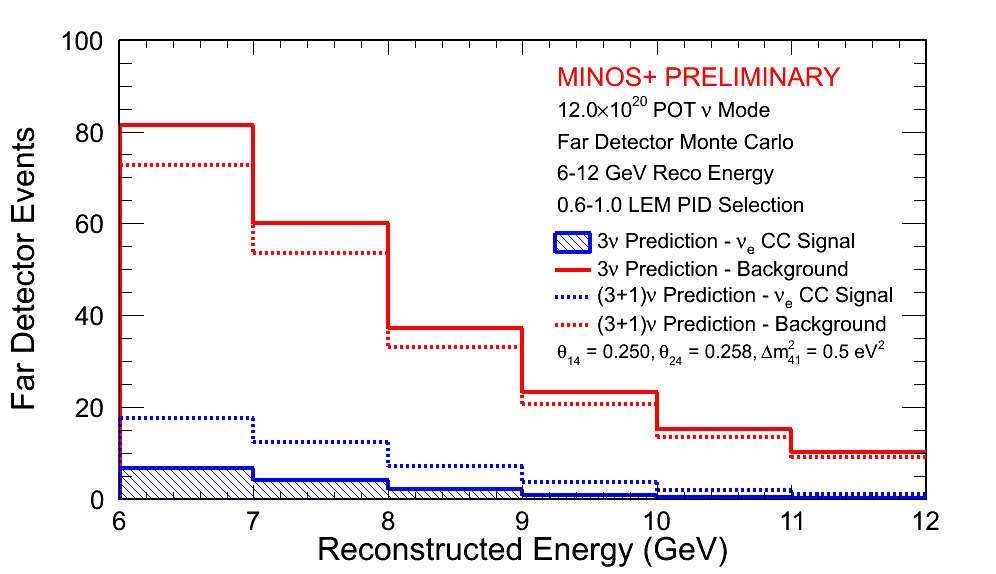}
\caption{Monte Carlo Signal and Background reconstructed energy distributions. Solid lines correspond to 3$\nu$ model predictions, while the dashed lines correspond  (3+1)$\nu$ model predictions for the given values of parameters.}
\label{fig:RecoE}
\end{figure}

The signature of $\nu_\mu \to \nu_s$ oscillations in MINOS manifests itself as an excess of $\nu_e$-CC events in the Far Detector, depending upon the particular $\Delta m^2_{41} $ and neutrino energy at the same time causing energy-dependent depletion in both NC and $\nu_\mu$-CC energy spectra. This is demonstrated in Figure~\ref{fig:RecoE} for the shown parameters. 
%
%
Signal and Background Monte Carlo distributions show the excess in the signal and depletion in the background energy spectra in the presence of nonzero $\theta_{14}$  and  $\theta_{24}$  mixing angles.

\begin{figure}[]
\centering
\includegraphics[height=3.1in]{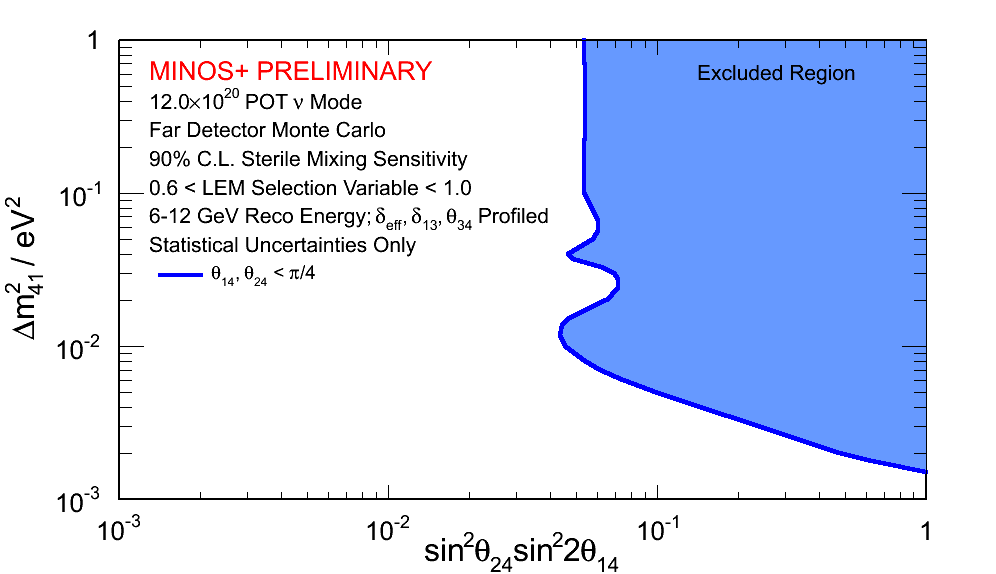}
\caption{90\% confidence level contour shows the MINOS+ allowed region in the SBL parameter space, assuming both 
$\theta_{14}$, $\theta_{24} < 0.25\pi $. }
\label{fig:sens1D}
\end{figure}

The sensitivity plots were generated by applying Gaussian intervals to the log-likelihood surfaces produced for specified values of 
$\Delta m^2_{41}$. At each $\Delta m^2_{41}$ value, a new surface was produced by profiling the combinations of $\theta_{34}$, $\delta_{CP}$, and $\delta_{eff}$.
Figure~\ref{fig:sens1D} shows the 90\% confidence level contour for MINOS+  in the SBL parameter space. The shaded blue region is excluded MINOS+.
  
\begin{figure}[]
\centering
\includegraphics[height=3.1in]{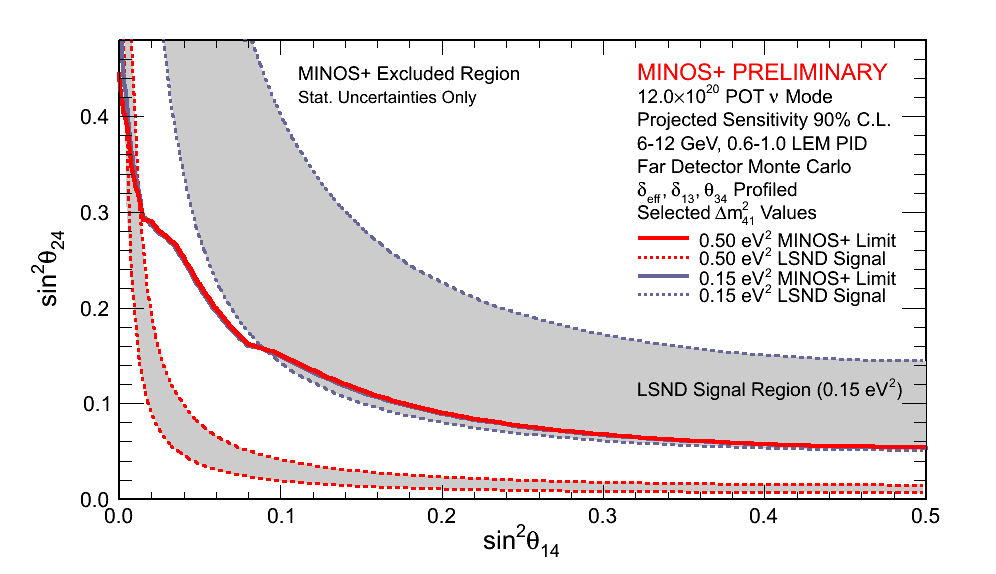}
\caption{90\% C.L. in two-dimensional parameter space for two chosen values of $\Delta m^2_{41}$, assuming $\theta_{14}$, $\theta_{24} < 0.25\pi $. MINOS+ excludes the parameter space to the upper-right of the solid contours. The shaded bands represent the LSND signal region associated with the indicated values of  $\Delta m^2_{41}$.}
\label{fig:sens2D}
\end{figure}

The MINOS+ experiment is capable of decoupling the  $\theta_{14}$ and $\theta_{24}$. In Figure~\ref{fig:sens2D} the shaded regions correspond to LSND  allowed regions. The blue dashed curves show  the $\Delta m^2_{41} = 0.15 eV^2$ LSND region, while the red dashed lines outline the $\Delta m^2_{41} = 0.50 eV^2$ LSND region. The solid lines show the MINOS+ 90\% C.L. contours for the same two values of  $\Delta m^2_{41}$. MINOS+ excludes the parameter space to the upper-right of the solid contours. For $\Delta m^2_{41} = 0.15 eV^2$ MINOS+ can exclude most of the LSND signal region, and although for $\Delta m^2_{41} = 0.50 eV^2$ MINOS's sensitivity is relatively poor, it still can exclude some portions of the LSND signal parameter space. 
Higher values of $\Delta m^2_{41}  (1-10^2 eV^2)$ can also be probed, but this will require incorporating oscillations in the Near Detector when extrapolating  to Far Detector.

\section{Summary}
In conclusion, MINOS+ is collecting new high-statistics data with medium energy beam. This will allow new precision measurements of 3-flavor oscillation parameters in an unexplored energy range.  It opens pathway to probe sterile neutrino oscillation search in $\nu_e$ appearance channel as well as  explore non-standard neutrino interaction scenarios in the 6-12 GeV region.
MINOS+ $\nu_e$ appearance can be used to set new constraints on low-mass sterile mixing parameters.

\Acknowledgments

This work was supported by the U.S. DOE; the United Kingdom STFC; the U.S. NSF; the State and University of Minnesota; Brazil's FAPESP, CNPq and CAPES. We are grateful to the Minnesota Department of Natural Resources and the personnel of the Soudan Laboratory and Fermilab for their contributions to the experiment.


\begin{thebibliography}{99}
\bibitem{lsnd}
LSND Collaboration: A.Aguilar et al., Evidence for neutrino oscillations from the observation of $\nu_e$ appearance in a $\nu_\mu$ beam, Phys. Rev. {\bf D64}, 112007 (2001).

\bibitem{miniboone}
 MiniBooNE Collaboration: A.A. Aguilar-Arevalo et al., Improved Search for $\nu_\mu \to \nu_e$ Oscillations in the MiniBooNE Experiment, Phys. Rev. Lett. {\bf 110}, 161801 (2013).

\bibitem{minos}
MINOS Collaboration: D.G Michael et al., The magnetized steel and scintillator calorimeters of the MINOS experiment, Nucl. Instrum. Meth. {\bf A 596}, 190 (2008).



\end{thebibliography}
\end{document}